\documentstyle[12pt]{article}
\newcommand{\ber}{\begin{eqnarray}}
\newcommand{\eer}{\end{eqnarray}}
\newcommand{\bea}{\begin{equation}}
\newcommand{\eea}{\end{equation}}
\begin{document}
\title{Evolution: The Case of the\\ 
Glyceraldehyde-3-Phosphate \\
Dehydrogenase Gene} 
\author{$Sujay\:Chattopadhyay^{1},\:Satyabrata\:Sahoo^{2},\:Jayprokas\:Chakrabarti^{1,\star}$\\\\
$^1$ Department of Theoretical Physics, Indian Association for the Cultivation of \\
Science, Calcutta 700 032, INDIA. \\
$^2$ Institute of Atomic and Molecular Sciences, Academia Sinica, \\
P.O. Box 23-166, Taipei, Taiwan 10764, Republic of CHINA. \\\\
$^{\star}$ To whom correspondence should be addressed. \\
E-mail: tpjc@mahendra.iacs.res.in \\
Telephone: 91-33-473 4971 / 3372 / 3073 extn. 286 \\
Fax: 91-33-473 2805}
\date{}
\maketitle
\newpage
\begin{abstract}
The enzyme Glyceraldehyde-3-Phosphate Dehydrogenase (GAPDH) catalyses the decomposition of glucose. The gene that produces the GAPDH is therefore present in a wide class of organisms. We show that for this gene the average value of the fluctuations in nucleotide distribution in the codons, normalized to strand bias, provides a reasonable measure of how the gene has evolved in time.
\end{abstract}
\newpage
$\bf{\:Introduction}$
\\
Evolution makes lower organisms into higher ones. The distribution of the nucleotides in the genes that code for proteins undergo changes in the process. It is sometimes assumed these variations in the nucleotide distributions come about due to random mutations. In this work we present quantitative evidence that the changes in the bases of the GAPDH are remarkably well ordered.
\par
The DNA sequence that codes for a single protein evolves as we go from one organism to the next. The evolution of the base composition of A, T, G and C for the same protein is the key to the dynamics of biological evolution. Some proteins are restricted to few organisms, others are more common. Amongst these proteins / enzymes, the glyceraldeyde-3-phosphate dehydrogenase (GAPDH) is present in all living organisms, as the key enzyme in glycolysis, the common pathway both in organisms that live in free oxygen and the ones that do not. The GAPDH catalyzes the dehydrogenation and phosphorylation of glyceraldehyde-3-phosphate to form 1,3-bisphosphoglycerate.
\par
The nature of the base organisation of the DNA sequences has been studied in the recent years (Voss 1992; Li and Kaneko 1992; Peng et al. 1992). The fractal correlations of $\frac{1}{f^\beta}$ type have been reported. These fractal correlations are more pronounced for the introns and the intergenic flanks. The exons, on the other hand, are characterised by strong peak at f=$\frac{1}{3}$ in the power spectrum. Here we work only with the exon regions and attempt to isolate the physical quantity that provides insights into the nature of evolution in the GAPDH.  
\par  
With this in mind we pick the DNA sequences coding for the GAPDH enzyme from a wide variety of prokaryotes, that include both bacteria and archaea (Woese et al. 1990), and eukaryotes (organisms with nucleated cells). Bacteria, in our study, is again subdivided into three groups: proteobacteria, $\it{Bacillus}$/$\it{Clostridium}$ group and cyanobacteria. Due to paucity of data for archaeal GAPDH, we cannot subdivide the archaea; we compare it as a whole with the groups of bacteria under Prokaryota. 
\par
Zuckerkandl and Pauling (1965) laid the basis for the study of genes and proteins for evolution. Over the years there have been the search for the universal common ancestor (Volkenstein 1994; Doolittle and Brown 1994; Woese 1998; Doolittle 1999; Woese 2000; Doolittle 2000) that may have preceded the prokaryotes and the eukaryotes. The studies on the ribosomal RNA provided some of the insights (Woese and Fox 1977a, 1977b; Fox et al. 1980). The relative importance of the elements, such as mutations, lateral gene transfer (Krishnapillai 1996; Brown and Doolittle 1997; Jain et al. 1999; Ochman 2000), that drive the evolution of species continues to be under active investigation. In our work here with the GAPDH we try to isolate the physical quantity (called X) that measures the evolution in this gene.\\ 
\newpage\noindent
$\bf{\:Number\:Fluctuations}$
\\
The coding sequences of the GAPDH genes from 42 different species, with 31 eukaryotes and 11 prokaryotes, were chosen (Source: GenBank and EMBL nucleotide sequence databases). These sequences have different distribution of the bases A, T, G and C. Since the codons are made of 3 of these bases, we divide the sequence into codons, i.e. choose the window size 3 bases long.
\par
On these windows of size 3, we compute the square of the numbers of A, T, G and C and define N(3) as: 
\ber
N(3)\:=\:n_{A}^{2}(3)+n_{T}^{2}(3)+n_{G}^{2}(3)+n_{C}^{2}(3)\nonumber
\eer
where $n_{A}^{2}(3)$,$n_{T}^{2}(3)$,$n_{G}^{2}(3)$,$n_{C}^{2}(3)$ are the numbers of A, T, G and C respectively in the codon window of size 3. Thus if, for instance, A occurs in all the three positions we get N(3)=9. If two are identical we get N(3)=4+1=5. If all the positions are occupied by different nucleotides, we get N(3)=1+1+1=3.
\par 
Thus N(3), for the window size 3, varies from 3 to 9 as we go from one codon to the next along the gene. We then compute the average value of N(3), call it $<N(3)>$, over the sequence. We notice here that a high value of $<N(3)>$ implies repeats of the bases. This means persistent sort of correlation amongst the bases. In other words, higher value of $<N(3)>$ implies a higher probability that the A, for instance, is going to be followed by the A. Conversely a lower value of $<N(3)>$ implies an antipersistent order in the sequence leading to a lower probability for the A to be followed immediately by the A.
\par
What do we expect for $<N(3)>$ for the random sequence of identical strand bias? Strand bias is the proportion of A, T, G and C in the sequence. These proportions vary as we go from one GAPDH sequence to another. We want to isolate the effect above and beyond the strand bias, therefore, study the quantity X defined as:
\bea
      X\:=\:\frac{<N(3)>}{<N(3,r)>} 
\eea
where $<N(3,r)>$ is the average value of the quantity $N(3)$ for the random sequence of identical total length and strand bias.
\par
$<N(3)>$ is measured for the sequences, while $<N(3,r)>$ is calculated using a 4-dimensional walk (Montroll and West 1979; Montroll and Shlesinger 1984) model. Hence the quantity X is obtained.
\par
To calculate $<N(3,r)>$ consider the following walk model in 4-dimensions corresponding to A, T, G and C. If we encounter the symbol i (i=A, T, G and C) we move one step along i. In this directed walk the probability function for a single step clearly is :
\bea
    P_{1}(x)\:=\:\sum_{i}p_i\delta(x_i-1)
\eea
where x$\equiv$(x$_A$,x$_T$,x$_G$,x$_C$), and p$_i$=$\frac{n_i}{N}$; n$_i$ is the number of times the symbol i appears in the sequence; N is the total number of symbols, i.e. the length of the sequence. We want to get the distributions after m steps, and therefore, define the characteristic function of the single step:          
\bea
\tilde P_1(k)\:=\:\sum_{i}p_{i}e^{ik_i}.
\eea
For m steps:
\bea
\tilde P_m(k)\:=\:[\sum_{i}p_{i}e^{ik_i}]^m
\eea
The quantity m is clearly the total number of steps, i.e. the window size. The moments of the distribution may be obtained by differentiating $\tilde P_m(k)$ with respect to k. In particular $<N(3,r)>$ is just the second moment of distribution and obtained from $\tilde P_m(k)$:
\bea
<N(3,r)>\:=\:[\sum_{i}\frac{\partial^2{\tilde P}_m(k)}{\partial k_i^2}]_{k_i \to 0}
\eea
Using (4) and (5), we get:
\bea
<N(3,r)>\:=\:m[(m-1)\sum p_i^2 + 1]
\eea
where we have used the relation $\sum p_i=1$.
\par
To crosscheck this relation, let us first set p$_A$=1; p$_T$=p$_G$=p$_C$=0. This is the case of maximal persistence. All the three bases, in this limit, are identical. From (6), we find:
\bea
<N(3,r)>\:=\:9, 
\eea
as we expect.
\par
To check again set p$_A$=p$_T$=p$_G$=p$_C$=$\frac{1}{4}$. The average value, from (6), gives:
\bea
<N(3,r)>\:=\:4.5 
\eea
For the window size m=3 the possible choices consistent with p$_A$=p$_T$=p$_G$=p$_C$=$\frac{1}{4}$ are 4x4x4=64, namely, the 61 codons + 3 stop codons. Calculation of the $<N(3,r)>$ for these 64 combinations is straightforward and gives the value 4.5 in agreement with (8).
\\\\
$\bf{\:Nucleotide\:Sequence\:Comparison}$
\\
The pairwise sequence alignment tool (ALIGN at the Genestream network server) available in the public domain gives a measure of the $\lq$$\lq$distance" (or the cross correlations) between the sequences. These distances provide additional data towards the study of evolution in the GAPDH gene.
\par
In the usual studies of evolution and phylogeny one relies exclusively on nucleotide sequence comparison. The rules used for alignment of sequences are constructed to give rise to the known pattern.
\par
In contrast, the change in the value of the X appears to us as the physical quantity of interest in the evolution in the GAPDH gene. The nucleotide sequence comparison we use in this work as supplementary, supportive data. \\
\par\noindent
$\bf{\:The\:X\:of\:Evolution}$ 
\\
The X values for the eukaryotes and the prokaryotes, for the GAPDH, for window size of 3, are given in Table 1. 
\par
Interestingly, the table 1 suggests two parallel lines of evolution, one for the prokaryotes; the other for the eukaryotes. Note the value of the X for the cyanobacterial genes is closer to that for the amphibian gene. The values for $\it{Bacillus}$/$\it{Clostridium}$ group and archaea are more or less the same as those for fish, and higher invertebrates such as arthropods.  
\par
As we look separately amongst the prokaryotes and the eukaryotes the X values increase as follows:\\
Prokaryota: $proteobacteria<archaea<\it{Bacillus}/\it{Clostridium}\: group<cyanobacteria$\\
Eukaryota: $fungus<invertebrate<fish<amphibia<bird<mammal\: (excl.\:human)<human$  
\\
It is to be remembered that in arriving at this increasing pattern the average value of the X over the members of the group has been considered. Within each group there are variations in the X (see Table 1).
\par
Assume now the GAPDH gene began from common universal ancestor. The route diverged to give proteobacteria on one side; fungal and invertebrate genes on the other. The proteobacterial gene develops further into three, archaeal, $\it{Bacillus}$/$\it{Clostridium}$ group and cyanobacterial, genes. The other trail from the fungus goes through fish, amphibia, probably reptilia for which the data is unavailable, birds and other mammals to reach its peak on humans. 
\par
Some groups have hypothesized that the eukaryotic species originated as the archaeal (e.g. $\it{Thermoplasma}$-like organisms) and the bacterial (e.g. $\it{Spirochaeta}$-like organisms) cells merged in anaerobic symbiosis and the GAPDH gene was contributed by the bacterial partner (Martin et al. 1993; Margulis 1996). Our results do not disprove this assumption. The X value averaged over all members of bacteria (i.e. proteobacteria + $\it{Bacillus}$/$\it{Clostridium}$ group + cyanobacteria) becomes 0.9662 $\pm{0.028}$ that is close to the X values for the invertebrates and the fungi (Table 1). 
\\
\par\noindent          
$\bf{\:Sequence\:Comparison}$ 
\\
The pairwise alignment tool gives a measure of similarity, or distance, between the various GAPDH genes under consideration (Figure 1).
\par
The results are fairly consistent with the picture that emerges from the study of the X. It suggests that the eukaryotic GAPDH genes might have originated from some eubacterial genes (Martin et al. 1993; Margulis 1996). 
\par
The alignment tool also suggests that both archaea and cyanobacteria may be quite distant from all other groups (Hensel et al. 1989; Arcari et al. 1993). As we measure the sequence similarity of the archaeal and the cyanobacterial genes with genes from the other two prokaryotic groups, we find the $\it{Bacillus}$/$\it{Clostridium}$ group gene closer to them than the proteobacterial one. This too supports the view obtained from the X values of the prokaryotes.    
\\
\par\noindent
$\bf{\:The\:X\:Evolution\:of\:the\:GAPDH\:Exon}$ 
\\ 
The plot of X for eukaryotes against their approximate period of origin in the geological time scale (Table 2) gives a fairly linear fit. We try a fit of the form $y=Kx+c$. For the slope $K$ for the eukaryotes we get: 
\bea
K_{euk}\:=\:\frac{\Delta X}{\Delta T}\:=\:1.1\:\times\:10^{-4}\:(\pm 0.2\:\times\:10^{-4})\:\:\:(myr)^{-1}, 
\eea
where myr$\equiv$million years. The computed $\chi^2$ value is 0.00009 with 6 degrees of freedom.\\ 
The earliest lifeforms are thought to come about around 3500 million years before present (myr BP). Presently we presume them as the proteobacterial ones. If the slope of the prokaryotic GAPDH gene X-evolution is assumed close to that for the eukaryotes, (9), then the cyanobacteria must have arisen
\bea
\Delta T\:=\:K^{-1}_{euk}\:[X_{cyano}\:-\:X_{proteo}]\:=\:493.5\:(\pm 126.6)\:\:\:(myr)
\eea         
after the proteobacteria. In myr BP this is 3500 - [493.5 ($\pm$126.6)] = 3006.5 ($\pm$126.6). Similarly, the periods of origin of the $\it{Bacillus}$/$\it{Clostridium}$ group and the archaea may be arrived at, and given in Table 3 and Figure 2. 
\par
Fossil stromatolites are macroscopic structures produced by some species of cyanobacteria. These are believed to occur from the early Precambrian (i.e., 3000 myr BP) to the Recent period (Thain and Hickman 1994). This is in good agreement with (10) for the time of origin of cyanobacteria obtained from the X-evolution.
\par
For an alternate approach assume the cyanobacteria appeared around 3000 myr BP, and the proteobacteria 3500 myr BP. The rate of change of the X, i.e.
\bea
K_{pro}\:=\:\frac{X_{cyano}\:-\:X_{proteo}}{\Delta T}\:=\:1.05\:\times\:10^{-4}\:\:\:(myr\:BP)^{-1}
\eea
Thus the slope of the prokaryotic GAPDH gene X-evolution (11) comes out to be nearly identical to that for the eukaryotes (9).  Figure 3 shows the best linear fits for the prokaryotes and the eukaryotes, which appear as two almost parallel lines. \\\\
 
$\bf{\:Discussions}$
\\
For the GAPDH exon the quantity X rises uniformly on two almost parallel paths - one for the prokaryotes; the other for the eukaryotes. The uniformity of rise in the X with time implies the genetic evolution is well-ordered; not the result of some random mutations.
\par
The rise of the X implies the trend towards persistent correlations in the base arrangement of codons. That is, as we go up the ladder of evolution the probability that a nucleotide, for instance the A is followed by the A increases. Note the result is true for the window of size 3. Whether the increase in persistence continues for any window size remains outside the scope of our analysis. The increase in persistence in the window of size 3 gives a measure of the complexity of the sequences at this scale (Rom{\'a}n-Rold{\'a}n et al. 1998). The diffusive processes that have persistence are being studied widely in recent years. For the GAPDH gene, suppose we work in the basis of purine-pyrimidine instead of the full A, T, G and C. We find, amusingly, the persistent nature of the diffusion increases even more for the window of size 3. Going beyond the GAPDH we find there are other important genes that share these features. 
\par
For the archaea the sequence comparisons indicate that they are more or less equally distant from the other prokaryotes and the eukaryotes. Yet the X-measure of the archaea places them between the proteobacteria and the $\it{Bacillus}$/$\it{Clostridium}$ group. The sequence information for the vertebrate GAPDH genes, especially for the amphibia, as of now, is limited. The availability of more data would improve the results to a considerable extent.     
\par
The ordered, uniform X-evolution of the GAPDH exon allows us to estimate the times of origins of $\it{Bacillus}$/$\it{Clostridium}$ group, cyanobacteria, archaea. The time of origin of cyanobacteria falls near the previous estimates.
\par
To conclude, the GAPDH gene is shown to be a marker for evolution. Importantly, the physical quantity X, the second moment of the codon base distribution, normalised to the strand bias, bears the footprint of a remarkably ordered evolution. 
\\\\
$\it{Acknowledgments}$. We thank Prof. S. Dey of Biotechnology Centre, IIT, Kharagpur, and Prof. Anjali Mookerjee of Sivatosh Mookerjee Science Centre, Calcutta, for discussions. Anup Som, our companion in the laboratory, has helped us in many ways.
\newpage

$\bf{References}$\\\\
ALIGN at the Genestream network server, Institut de G{\'e}n{\'e}tique Humaine, Montpellier, France (http://www2.igh.cnrs.fr/bin/align-guess.cgi) \\
Arcari P, Russo AD, Ianiciello G, Gallo M, Bocchini V (1993) Nucleotide sequence and molecular evolution of the gene coding for glyceraldehyde-3-phosphate dehydrogenase in the thermoacidophilic archaebacterium Sulfolobus solfataricus. Biochem Genet 31:241-251 \\
Brown JR, Doolittle WF (1997) $\it{Archaea}$ and the Prokaryote-to-Eukaryote transition. Microbiol Mol Biol Rev 61:456-502 \\
Doolittle WF, Brown JR (1994) Tempo, mode, the progenote, and the universal root. Proc Natl Acad Sci USA 91:6721-6728 \\
Doolittle WF (1999) Phylogenetic classification and the universal tree. Science 284:2124-2128 \\
Doolittle WF (2000) The nature of universal ancestor and the evolution of proteome. Curr Opi Struc Biol 10:355-358 \\
Fox GE, Stackebrandt E, Hespell RB, Gibson J, Maniloff J, Dyer TA, Wolfe RS, Balch WE, Tanner RS, Magrum LJ, Zablen LB, Blakemore R, Gupta R, Bonen L, Lewis BJ, Stahl DA, Luehrsen KR, Chen KN, Woese CR (1980) The phylogeny of prokaryotes. Science 209:457-463 \\
Hensel R, Zwickl P, Fabry S, Lang J, Palm P (1989) Sequence comparison of glyceraldehyde-3-phosphate dehydrogenases from the three urkingdoms: evolutionary implication. Can J Microbiol 35:81-85 \\ 
Jain R, Rivera MC, Lake JA (1999) Horizontal gene transfer among genomes: The complexity hypothesis. Proc Natl Acad Sci USA 96:3801-3806 \\
Krishnapillai V (1996) Horizontal gene transfer. J Genet 75:219-232 \\ 
Li W, Kaneko K (1992) Long range correlation and partial 1/$f^{\alpha}$ spectrum in a noncoding DNA sequence. Europhys Lett 17:655-660 \\
Margulis L (1996) Archaeal-eubacterial mergers in the origin of Eukarya: Phylogenetic classification of life. Proc Natl Acad Sci USA 93:1071-1076 \\ 
Martin W, Brinkmann H, Savonna C, Cerff R (1993) Evidence for a chimeric nature of nuclear genomes: eubacterial origin of eukaryotic glyceraldehyde-3-phosphate dehydrogenase genes. Proc Natl Acad Sci USA 90:8692-8696 \\ 
Montroll EW, West BJ (1979) In: Montroll EW, Lebowitz JL (eds) Fluctuation Phenomena. North-Holland, Amsterdam \\   
Montroll EW, Shlesinger MF (1984) In: Lebowitz JL, Montroll EW (eds) Nonequilibrium Phenomena II From Stochastics to Hydrodynamics. North-Holland, Amsterdam \\
Ochman H, Lawrence JG, Groisman EA (2000) Lateral gene transfer and the nature of bacterial innovation. Nature 405:299-304 \\
Peng C-K, Buldyrev SV, Goldberger AL, Havlin S, Sciortino F, Simons M, Stanley HE (1992) Long-range correlations in nucleotide sequences. Nature 356:168-170 \\
Pough FH, Heiser JB, McFarland WN (1999) Vertebrate Life. Prentice-Hall, New Delhi \\
Rom{\'a}n-Rold{\'a}n R, Bernaola-Galv{\'a}n P, Oliver JL (1998) Sequence compositional complexity of DNA through an entropic segmentation method.
Phys Rev Lett 80:1344-1347 \\
Stein P, Rowe B (1995) Physical Anthropology. McGraw-Hill, Berkshire, UK \\
Thain M, Hickman M (1994) In: Thain M, Hickman M (eds) The Penguin Dictionary of Biology. Penguin Books, London, p 594 \\
Volkenstein MV (1994) In: Physical Approaches to Biological Evolution. Springer-Verlag, Berlin \\ 
Voss RF (1992) Evolution of long-range correlations and 1/$f$ noise in DNA base sequences. Phys Rev Lett 68:3805-3808 \\
Woese CR, Fox GE (1977) Phylogenetic structure of the prokaryotic domain: the primary kingdoms. Proc Natl Acad Sci USA 51:221-271 \\
Woese CR, Fox GE (1977) The concept of cellular evolution. J Mol Evol 10:1-6 \\ 
Woese CR, Kandler O, Wheelis ML (1990) Towards a natural system of organisms: Proposal for the domains Archaea, Bacteria, and Eukarya. Proc Natl Acad Sci USA 87:4576-4579 \\  
Woese CR (1998) The universal ancestor. Proc Natl Acad Sci USA 95:6854-6859 \\ 
Woese CR (2000) Interpreting the universal phylogenetic tree. Proc Natl Acad Sci USA 97:8392-8396 \\ 
Zuckerkandl E, Pauling L (1965) In: Bryson V, Vogel HJ (eds) Evolving Genes and Proteins. Academic Press, New York, pp 97-166 \\

\newpage\noindent
$ {\bf{Figure\:Legends}} $ \\\\
Figure 1. Average $\%$ identity of nucleotide sequence in the GAPDH genes from different groups of organisms. The black lines and values imply the alignment results between the proteobacterial gene and the genes from all other groups; the pink lines and values for the $\it{Bacillus}$/$\it{Clostridium}$ group gene with the other genes; the green lines and values between the archaeal gene and the other genes; and the blue lines and values for the cyanobacterial gene with the rest.
\\\\
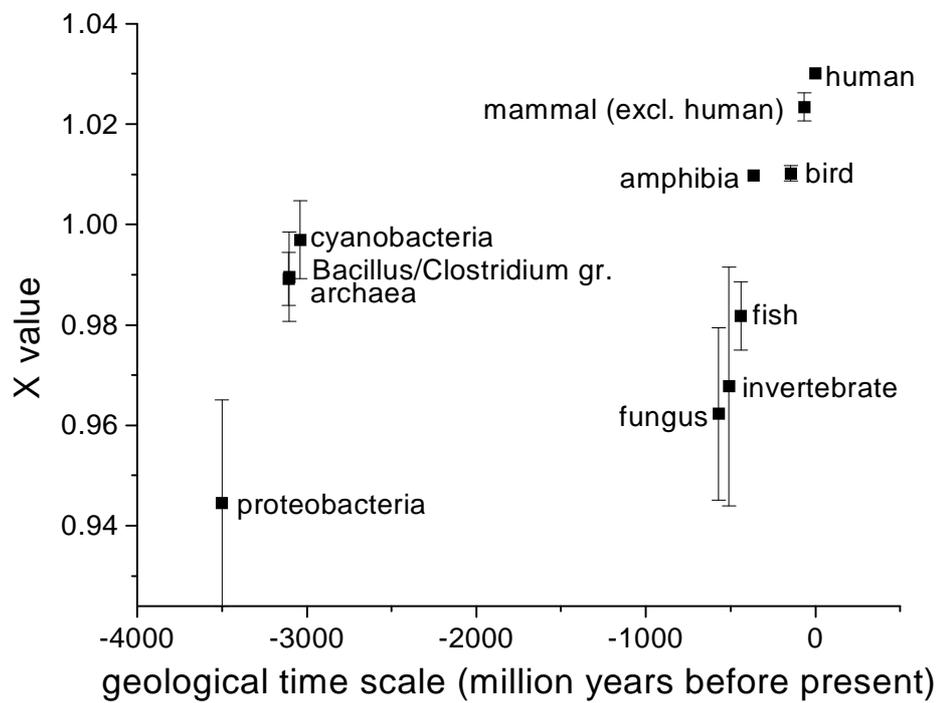
Figure 2. The probable periods of origin of the prokaryotes (see Table 3), along with the periods of origin of the eukaryotes (see Table 2), are plotted against the X values for the corresponding GAPDH genes (see Table 1). The error bars simply indicate the standard deviation from the average X values for the respective groups. Here the slope of the prokaryotic GAPDH gene X-evolution is assumed to be equal to that for the eukaryotes.
\\\\
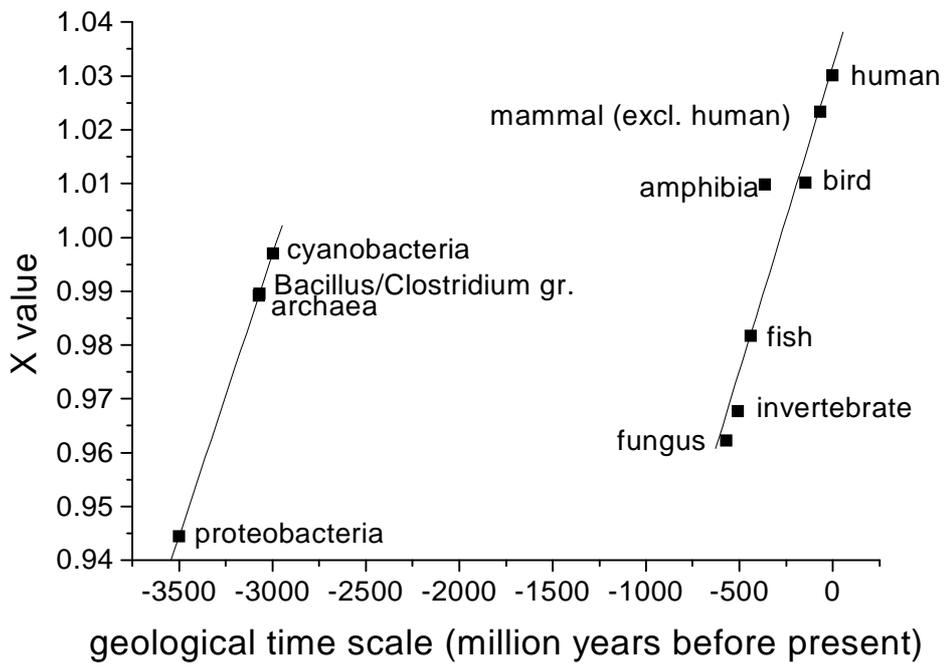
Figure 3. The best linear fit-curves both for the prokaryotes and for the eukaryotes, as we plot the X values vs. the periods of origin. The solid black lines denotes the best fit-curves. The slopes of the GAPDH gene X-evolution for the prokaryotes and the eukaryotes are found to be close enough to suggest two nearly parallel lines of evolution.

\newpage
\begin{table}
\renewcommand{\arraystretch}{1.5}
\begin{center}
\caption{\bf{The X values for prokaryotes and eukaryotes, along with the range of deviations in respective categories.}} 
\bigskip
\begin{tabular}{|c c|}
\hline
Category & X  \\
\hline
I. PROKARYOTA &      \\
$\:\:\:\:\:$ proteobacteria & 0.9445 ($\pm$0.0127) \\
$\:\:\:\:\:$ archaea & 0.9892 ($\pm$0.0075)  \\
$\:\:\:\:\:$ $\it{Bacillus}$/$\it{Clostridium}$ group & 0.9896 ($\pm$0.0126)  \\
$\:\:\:\:\:$ cyanobacteria & 0.9970 ($\pm$0.0110) \\
\hline
\end{tabular}
\end{center}
\renewcommand{\arraystretch}{1.5}
\end{table}

\newpage
\begin{table}
\renewcommand{\arraystretch}{1.5}
\begin{center}
\begin{tabular}{|c c|}
\hline
Category & X  \\
\hline
II. EUKARYOTA &      \\
$\:\:\:\:\:$ fungus & 0.9623 ($\pm$0.0121) \\
$\:\:\:\:\:$ invertebrate & 0.9677 ($\pm$0.0134)\\
$\:\:\:\:\:$ fish & 0.9819 ($\pm$0.0097) \\
$\:\:\:\:\:$ amphibia & 1.0098 \\
$\:\:\:\:\:$ bird & 1.0102 ($\pm$0.0021) \\
$\:\:\:\:\:$ mammal (excl. human) & 1.0234 ($\pm$0.0019) \\
$\:\:\:\:\:$ human & 1.0301 \\
\hline
\end{tabular}
\end{center}
\renewcommand{\arraystretch}{1.5}
\end{table}

\newpage
\begin{table}
\renewcommand{\arraystretch}{1.5}
\begin{center}
\caption{\bf{Origin of eukaryotes in geological time scale. }}
\bigskip
\begin{tabular}{|c c|}
\hline
Category & Position in time scale (myr BP) (Stein and Rowe 1995; Pough et al. 1999)  \\
\hline
 Fungus & 570 \\
 Invertebrate & 510 \\
 Fish & 439 \\
 Amphibia & 363 \\
 Bird & 146 \\
 Mammal (excl. human) & 66.4 \\
 Human & 1.64 \\
\hline
\end{tabular}
\end{center}
\renewcommand{\arraystretch}{1.5}
\end{table}

\newpage
\begin{table}
\renewcommand{\arraystretch}{1.5}
\begin{center}
\caption{\bf{Probable origin of prokaryotes in geological time scale as emerged from their X values.}}
\bigskip
\begin{tabular}{|c c|}
\hline
Category & Position in time scale (myr BP)  \\
\hline
 Proteobacteria & 3500 \\
 Archaea & 3079.5 ($\pm$108.2) \\
 $\it{Bacillus}$/$\it{Clostridium}$ group & 3076.0 ($\pm$108.9) \\
 Cyanobacteria & 3006.5 ($\pm$126.6) \\
\hline
\end{tabular}
\end{center}
\renewcommand{\arraystretch}{1.5}
\end{table}
\end{document}